# The Three Social Dimensions of Chatbot Technology

**Mauricio Figueroa-Torres[1]**

**Abstract**
The development and deployment of chatbot technology, while spanning decades and employing different techniques, require innovative frameworks to understand and interrogate their functionality and implications. A mere technocentric account of the evolution of chatbot technology does not fully illuminate how conversational systems are embedded in societal dynamics. This study presents a structured examination of chatbots across three societal dimensions, highlighting their roles as objects of scientific research, commercial instruments, and agents of intimate interaction. Through furnishing a dimensional framework for the evolution of conversational systems — from laboratories to marketplaces to private lives— this article contributes to the wider scholarly inquiry of chatbot technology and its impact in lived human experiences and dynamics.

## 1 Introduction

The global outlook on Artificial Intelligence (AI) has shifted substantially with the deployment and adoption of generative conversational systems. A 2023 survey, encompassing over 22,000 adults from 31 diverse countries, finds that two-thirds of respondents are convinced that AI will profoundly alter their lives in the near future (Stanford University, 2024). Chatbots are gaining ground in various areas of human life, from education to customer care, new trends in academic research, or the pursuit of emotional needs. Yet, these practices do not necessarily come together on a unified understanding of what chatbot technology actually does or how society engages with it. While chatbots are indeed algorithmic systems used to entertain

✉ Mauricio Figueroa-Torres
   m.figueroa-torres2@newcastle.ac.uk

[1]  Newcastle University, 19-24 Windsor Terrace, Newcastle Upon Tyne NE2 4HQ, UK

     ⌂ Springer

a conversational environment,[1] they influence differently three societal dimensions: scientific research, the market, and intimacy.

This multidimensional nature of chatbot technology is not merely semantic but indicative of varied lenses that illuminate different natures, dynamics, and experiences, what this article broadly refers to as *dimensions*: technological, commercial, and personal. This framework provides a more balanced examination and understanding of chatbots, explaining what brings them about and how they evolve, beyond their technological trajectory, in three identified dimensions.

This article is structured as follows. It begins by identifying and deconstructing the techno-centric narrative of chatbot development, positing that while technological milestones, such as advancements in natural language processing (NLP) and machine learning (ML), are undeniably essential, they do not necessarily capture the complex interplay of chatbots and human lived experiences. Subsequently, this article fleshes out each dimension and its characteristics.

In the first dimension, chatbots are as objects of scientific inquiry, driven by performance and computational capabilities. In the second dimension, chatbots become tools within commercial strategies, wherein businesses introduce chatbot technology in their operations and automate their transactions. In the third dimension, chatbots mutate into agents of social interaction, whereby users resort to conversational agents to satisfy specific emotional needs. While the dimensions may overlap, each of them offers a different set of characteristics on the role and nature of chatbots, thereby enriching our understanding and highlighting the multifaceted impacts of conversational systems. This differentiated exploration aims to provide a richer, more integrated understanding of chatbot technology, culminating in a discussion of the implications of these dimensions and setting a fertile groundwork for future academic and policy discussions.

## 2  Reassessing the Progression of Technology

This article joins the broader rejection of technological determinism. Technology does not possess "intrinsic independence from the social world", as it actually reflects the contexts and conflicts of its social adoption, making it a site of social scrutiny and critique (Feenberg, 1999). Along these lines, as Pinch and Bijker's contend, the way different *relevant social groups* relate to, or engage with, a given technological artefact — and the "meanings which those groups give to the artefact" — shape its development, underpinning the multidirectional character and its interpretative flexibility (Pinch & Bijker, 1984). But the same operates the other way around: artefacts change or alter society back. In Verbeek's words, it is a co-shaping dynamic whereby "human beings are not sovereign with respect to technology" but rather are inextricably interwoven with it' (Verbeek, 2005, p. 44) — hence technologies alter the human world as well.[2]

---

[1] While the field of computer science generally defines chatbots as computer programs designed to engage in human conversational settings, often through plain text, these systems employ a variety of techniques and are built upon diverse structural frameworks (McTear, 2020)— these will be discuss further in Section III.

Non-linear approaches expand academic inquiry and provide a more granular understanding of technological development. One certainly can find viewpoints that are grounded in similar approaches when it comes to interpreting technological and social dynamics. Floridi's work, when analyzing the evolution of information technologies, serves as an illustrative example of alternative framings of technological development. Rather than viewing information technologies as evolving in a straightforward progression from the techniques employed from Gutenberg to Turing, he emphasizes the growing dependency of human societies on digital tools—first for recording, then transmitting, and now processing information — enriching the importance of adopting varied interpretive frameworks to fully appreciate the complexities of technological advancement (Floridi, 2014).

Additionally, in technological realms other than digital technologies, such as vehicular evolution, the analysis may well transcend mere technical specifications such as engine power and design, and touch upon wider social and material structures (Perez, 2010), and with the diverse modalities — private, professional, shared, or collective — beyond their mechanical attributes and chronological progression.[3] In a similar vein, Tarleton Gillespie's observation on the multiple performability attributed to the term 'algorithm' illustrates the challenge of varied understanding and texture the very same word possess across different realms: technical experts, social scientists, and the general public (Tarleton Gillespie, 2016). The dimensional framework this article organizes around chatbot technology similarly acknowledges contextual flexibility.[4]

---

[2] As Floridi observes, every technology is 'designed according to some implicit or explicit values, by some people for some people, within a culture and with a culture in mind, for some uses rather than others…' (Floridi, 2023), yet I would add — in line with the aforementioned co-shaping optic — that every technology establishes its own practices and redefines human experiences. In that regard, Julie E. Cohen encapsulates this with remarkable precision, when she advances that 'In addition to considering whether technologies have politics, we must now consider whether (some) technologies leave room for humans to practice politics at all […] Even as we are busy configuring our tools, they are also busy configuring us.' (Cohen, 2017).

[3] Perez's analysis centers on what she terms *techno-economic paradigms*, and she clearly establishes that each of the five technological revolutions she identifies — Industrial; Steam & Railways; Steel & Electricity; Oil, Automobile & Mass Production; and Information & Telecommunications — consists of clusters of innovations. In her framework, conversational AI, generative AI, and arguably AI at large fall within the ICT revolution rather than representing a new, separate revolution. Her analysis, therefore, is useful in understanding how certain inventions within each revolution extend beyond their technical functionalities and original conceptions, influencing and shaping societies on different levels: 'When they are sufficiently radical, innovations stimulate whole industries. Thus the emergence of television led to the growth of industries that manufacture receiving and broadcasting equipment, as well as of multiple specialised supplier industries. At the same time it spurred the transformation of the producing and advertising industries, film, music and other creative sectors, plus new maintenance and distribution activities and so on […] new technology systems not only modify the business space, but also the institutional context and even the culture in which they occur.' (Perez, 2010, p. 188).

[4] As a matter of fact, even the term chatbots may carry diverse understandings across different settings: for computer scientists, these represent technological systems subject to computational capacity; in the

Stated differently, chronological analysis clarifies the sequence of innovations, but it falls short of explaining how society shapes technology and how technology, in turn, reshapes society— in clear contrast to the techno-determinist paradigm.

Thus, the societal dimensions of conversational systems mold, and are molded by, different elements. Although sequential milestones contribute to our understanding, there are different narratives, experiences, and agendas that endow chatbot technology — and even the term itself — with varying impact and roles across different dimensions. In the following section, I will clarify the rationale behind focusing on three dimensions — scientific research, market practices, and personal intimacy — as central to this analysis, and why the term *dimension* is preferred. The mapping of the societal dimensions of chatbot technology is not presented as a definitive or ultimate categorization, but rather a contribution that helps to structure sociotechnical critiques. With this in mind, the present analysis takes place.

The ensuing section will provide a brief overview of progression of chatbot technology around its technological affordances, highlighting key milestones, inventions, and advancements in a temporal order. In the subsequent section I will reassess this through the lens of the social dimensions that conversational systems permeate and influence.

## 3 The Techno-Centric Narrative of the Evolution of Chatbots: Significance and Omissions

While this article advocates for a dimensional understanding of chatbot technology, it does not deny the role historical progression plays in elucidating the functional mechanics and capabilities of conversational systems. The purpose of this section is to trace the evolution of key milestones in conversational AI, to later compare it with the societal dimensions that will be articulated later in this article.

However, three important caveats must be made.

First, it is crucial to acknowledge that early scholarship critically engaged with conversational AI and placed significant emphasis on the societal implications of these technologies.[5]

Second, the valuable scholarship engaging with societal implications of conversational AI must be underscored, and the intent of this article is precisely to build upon and reorganize those foundations, rather than to dismiss or omit them. In fact,

---

Footnote 4 (continued)

commercial domain, they are seen as opportunities for growth and efficiency; for end-users, they signify the introduction of novel interactive experiences.

[5] However, these inquiries have found placed into separate academic domains apart from the engineering and computational theory that prioritize technical design and optimization. This is manifest, for instance, in the separation of academic departments where sociotechnical critique falls outside the scope of computer science departments and schools (Stanford, 2024; UCL, 2024), the sources of funding for projects, different employment sectors and recruitment channels, and the relatively low presence of sociotechnical scholarship in major conversational AI conferences and computers science more broadly (*CUI Conference—Proceedings*, 2024; NeurIPS, 2024).

the value and existence of critical scholarship is not in question, but rather this paper is concerned in constructing a framework that helps to understand how chatbot technology has unfolds differently across three societal dimensions.

Thirdly, this section is not meant to serve as a comprehensive review of all relevant and technical computer science literature but rather to frame the technocentric lens that spreads through much of the discourse on chatbot evolution and discusses it from a technical standpoint.

Having clarified the aforementioned points, the next paragraphs will provide a brief account of the technological and linear progression of conversational AI.

Alan Turing's foundational contributions to computational theory, particularly as articulated in his post-war paper on machine intelligence (Turing, 1950), set the stage for computational coding and a new wave of experimentation. Turing's endeavors reflect the computational challenge to make machines *think* and, fifteen years later, Joseph Weizenbaum followed that path—attempting to make machines *talk*.

In 1966, Weizenbaum's ELIZA captivated the scientific community at the Massachusetts Institute of Technology (MIT) and beyond. It was a rules-based system to make a computer simulate a psychotherapist to entertain a human-like conversation (Weizenbaum, 1966). As I will analyze in the following section, the unprecedented social effects and user engagements with and around ELIZA turned Weizenbaum himself a strong critic of the adoption of AI and the potential dependency on these systems (Weizenbaum, 1976). But this did not digress or stop the continuous experimentation in the field.

Despite early critiques of conversational AI and its societal implications, the timeline continues with further experiments building upon the technological foundation set by ELIZA, laying the groundwork for the chronological progression of conversational AI, such as PARRY in 1972, designed to simulate a patient with schizophrenia, thus reflecting the counterpart to Freud's divan (Adamopoulou & Moussiades, 2020). The chronology extends into periods of reduced research activity, often referred to as the AI Winter, which was marked by a notable decline in field in terms of funding and research (Schuchmann, 2019). Within this technological progression, key milestones took place at the end of the twentieth century. Noteworthy is the creation of *Chatterbot* by Michael Mauldin in 1994, a conversational system intended to interact with players in a video game setting. This system built upon the basic conditional logic structures of ELIZA and PARRY and ultimately led to the popularization of the term 'chatbot' (Mauldin, 1994). The technological account continues with the availability of widespread Internet access, and how it further facilitated the development of conversational systems through remote interaction and personal computers, like the successful web-deployed conversational system ALICE in 2001 (AbuShawar & Atwell, 2015).

The chronological progression of chatbots is, therefore, marked by predefined response sets, memory improvements, keyword matching, but also by the revival of techniques, once restricted by technological limitations, such as poor processing power and inadequate datasets, that later redefined the field of conversational AI. The introduction of machine learning (ML) techniques marked a turning point, transforming chatbots into more sophisticated systems capable of generating spontaneous, human-like responses. An early example of this shift is Rollo Carpenter's Cleverbot, launched in 2008, employing pattern recognition techniques to generate innovative, spontaneous replies (Fryer et al., 2020). Subsequent research illuminated the technological progression and potential to harness larger datasets, as demonstrated in the pioneering study by Vinyals and Le in 2015, which involved designing a chatbot that generated responses based on movie subtitles databases (Vinyals & Le, 2015).

From a technical standpoint, the transition from rule-based systems to ML models marked a crucial advancement in the evolution of conversational AI. However, a new frontier of this field has been defined by emerging methodologies such as large-scale web scraping, the integration of vast data sets, and the leveraging of increasingly robust computational architectures. The generation and application of large language models (LLMs)—which represent a significant subset of current ML research—constitute a new paradigm in the technological trajectory of conversational agents— including coding, writing auto-completion, grammar assistance, fueling search engine responses, answering questions, and overall increasing the difficulty of distinguishing synthetic text from human-written text — adding a further layer of complexity to machine-mediated dialogue (Brown et al., 2020). This progression has also sparked critical discussions on the technical limitations and flaws, exemplified by Emily Bender's et al. coining of the term *stochastic parrots* to describe the superficiality of these models by merely predicting sequential word probabilities (Bender et al., 2021). Additionally, a significant body of scholarly work has emerged to scrutinize the shortcomings and characteristics of these systems, emphasizing their 'foundational' yet incomplete nature and their potential adaptability across various domains (Center for Research on Foundation Models, 2021).

The technological trajectory from basic programs like ELIZA to the latest LLM-enabled chatbots is indeed a clear reflection of growing computational sophistication and the ability to simulate increasingly complex human-like conversations. But these developments also engage with and transform various aspects of human interaction and societal structures. The dimensional understanding of chatbot technology requires embracing a sociotechnical approach to shed light on the evolution of chatbot technologies, the societal spaces they inhabit, and the human interactions and societal structures they touch.

Acknowledging the chronological and technological milestones of the field, the next section of this article constructs a non-linear interpretation of chatbot evolution. The intention or purpose is not to supplant the technological account but to broaden it, thereby enhancing our comprehension of how these systems influence and are integrated into human practices, giving rise to the three societal dimensions this article decodes.

## 4 Chatbots in Three Societal Dimensions

Chatbot technology evolves in three social dimensions: scientific research, the market, and intimacy. But what does this article mean when it talks about *dimensions*?

The choice of the term *dimension* is deliberate to indicate more than just phases, moments, or simply stages, which typically imply a chronological or linear sequence. Instead, dimensions are intended to acknowledge the complex patterns of influence, interconnected set of structures, and conditions of chatbot technology that bring them about in separate realms. Unlike phases or moments, dimensions can overlap and interact, thus reflecting the multi-layered ways in which chatbots influence and are influenced by their respective environments.

Each dimension has a different configuration and brings together a range of factors and elements. At times, the emphasis is closely related to *perception*, associated with how different actors — developers, intermediary providers, and customers— engage with chatbots. However, these dimensions are not limited to perception but extend to the understanding and broader impacts that chatbots have within specific contexts—whether in research labs, market dynamics, or end-user environments. The dimensions are not monolithic and have different characteristics. The dimensional framework is, therefore, an adaptable analytical tool, accounting for perception, patterns of influence, agendas, experiences, and broader technological impacts, that shifts depending on specific contexts.

Thus, this article does not reduce each dimension to a singular, uniform concept. Rather, it emphasizes the fluidity and context-dependent nature of these dimensions, understanding that technological phenomena —like chatbots—operate within nonuniform, overlapping landscapes.

## 4.1 Chatbots and Computer Labs: Chatbots as Objects of Scientific Research

Chatbots are indeed algorithmic systems. As such, there is one clear way to understand them: objects of computational capacity and experimentation. In this dimension, chatbots are niche experiments confined to a small number of labs. The dimension *chatbots as objects of scientific research* is the nascent circle of technology but clearly subsists to this day — it is therefore not a phase or a moment in the history of chatbot technology. I will now scrutinize what characterizes this dimension and how it overlaps with the other two. In this dimension, two primary elements are consistent: the determination of computer scientists to enhance the performance in key benchmarks of conversational AI, and the technological context that, initially, renders these technologies restricted and inaccessible to the general public.

For instance, by tracing back to the origins of computational linguistics, one may advert that the introduction of rule-based technology to enable computers mimic human conversation symbolises a crucial technological breakthrough. But, at the same time, this illustrates vividly the highly specialized sector involved and interested groups in the process. In that regard, landmarking moments like the seminal interaction with ELIZA responding 'In what way?' to the statement 'Men are all alike' (Weizenbaum, 1966, p. 36) does not only reveal the functionality of rule-based technology in simulating conversations, but it also depicts the highly specialized settings of early computational linguistics research, largely restricted to academic labs like MIT's. While naturally the landscape has changed dramatically from Weizenbaum's era to our time, the defining elements of this dimension are nonetheless continuous.

Early stages of digital technologies mostly involve technical problems that appeal mainly to specialists who have the expertise and motivation to engage with them. Hence, in this dimension, conversational systems are computer experiments not available to, and arguably not of the immediate interest of, the public. This echoes, for instance, with Latour's notion of transition in technology and science, where innovations pass through different phases and actors before they can be enrolled into a larger network that includes the wider public or *multitudes* (Latour, 1997). In other words, chatbots in this dimension are defined as technical experiments, and the *multitude* does not interact with or fully understand them because they remain within a limited network of experts.

The dimension of chatbots as objects of scientific research is continuous and non-dependent on a single set of techniques or functioning of the systems. I mentioned before that two defining elements are evident: the interest of computer scientists reflected in constant experimentation and the limited access to that technology. I will now proceed to explain this in further depth.

For instance, when a conversational system is envisioned, it is designed and tested in a confined computer environment, and some are never released to the public. While it is true that outputs and discussions around chatbot experiments are often shared and disseminated, these are usually in the form of scholarly articles, conference proceedings, and technical reports, which are not of interest, and mostly inaccessible, to the public. Consequently, some systems, while innovative, may remain just as lab experiments and findings may be shared within specific circles, like Chatterbot by Mauldin discussed in the previous section. Broader public awareness often coincides with a separate dimension that deals with the introduction to the market. Certain systems do transcend successfully to this second dimension and become public interfaces, like Cleverbot by Rollo or ALICE by Richard Wallace, also referred to in the previous section.

Indeed, the dimension of chatbots as objects of scientific research is far from static or monolithic. It has evolved in its composition and structure—hence another reason why the term *dimension* better reflects the inner interactions between actors, forces, dynamics, and perceptions. In the early days of computer science research, the military played a significant role (Birnhack & Elkin-Koren, 2003; Edwards, 1996). Within computational linguistics, it was primarily academics who shaped the development of conversational systems, establishing themselves as the primary architects shaping the trajectory of chatbot technology. The talent, resources, and expertise were concentrated within university settings. Interestingly, in the early days of chatbot research, Big Tech companies, as we know them today, played a minimal role in the field of computing research.

In contrast, today Big Tech companies have become the primary drivers of AI research. In this regard, the 2024 AI Index Report by Standford University highlights that in 2023, industry produced 51 notable ML models, while academia contributed only 15, and that by 2022 a significantly larger proportion of PhD graduates (70.7%) joined industry after graduation compared to those entering academia — in clear contrast, PhDs joining industry in 2011 were about 40.9% (Stanford University, 2024).

This data illustrates Big Tech's increasing investment and dedicated in-house technical teams enhancing model accuracy and algorithmic efficiency while exploring new applications across domains. And while academic institutions and scholars are certainly part of the equation, Big Tech companies take the lead in the agenda, in addition that they fund academic research, influencing the academic discussions, projects, and even the education landscape in the realm. Notwithstanding the changing texture of the 'chatbots as objects of scientific research' dimension, the interest of computer scientists to surpass the previous performance metrics in conversational AI, and the reduced access to the experimentation stage of these systems, prevail as defining elements of this dimension.

A closer look at the dynamics of scientific research in conversational AI further clarifies these two defining elements.

From Weizenbaum's early work to today, the design, architecture, and testing of these systems largely occur in closed, restricted environments—even with the rise of open-source initiatives and individual efforts, this is still a niche. Today, this process increasingly takes place within corporate settings, where much of the expertise driving the field now resides. Additional components have emerged in this dimension, such as the need for computer power, energy consumption or the capacity of servers, yet the lack of public awareness and engagement prevails. The lifecycle of AI products, particularly in conversational AI, underscores this defining component: the reduced access to the experimentation and creation of chatbots.

For example, OpenAI extensively worked on developing its capabilities in generative AI and LLMs well before offering a beta version to developers who, in turn, accessed the model via the API, and were allowed to "explore language modeling systems across a wide range of functions", with the aim to "better understand the behaviors, capabilities, biases, and constraints of large-scale language models" (OpenAI, 2020). Eventually, and following further finetuning and experimentation, it launched ChatGPT as an interface of its own to the general public (OpenAI, 2022), and "just three months" after its public release polls found "that 40% of respondents were using the tool in their professional lives"(Shevlin, 2024). Similarly, Google's Bard, later rebranded as Gemini, along with Meta's Llama and Anthropic's Claude, aligned to this pattern development and release characterized by an initial limited access where public knowledge is constrained. Therefore, this dimension underscores how chatbot advancements are fostered in highly insulated settings, curated by specialized teams dedicated to pushing the boundaries of AI capabilities, often operating under a veil of secrecy.

With regards to the interest of the scientific community, it should be highlighted that conversational AI researchers and scientists represent a diverse set of talent,[6] ranging from interns, PhD candidates, post-doctoral researchers, and so forth, in both corporate and academic environments, each bringing perspectives and motivations to their work: some may be focused on learning and experimentation, while others aim to expand the capabilities of theoretical knowledge or pursue long-term research agendas. In contrast, corporate leadership in Big Tech manifests in a more unified interests, such as profit maximization and competitive advantage, and product development. While the goals, interests, and visions of researchers and corporate heads can very well align, particularly in the pursuit of technological innovation, the dimensions they inhabit may differ. Thus, even though market forces play a crucial role, the impetus in experimentation and the technological advancement of the systems themselves in terms of capabilities, performance, and accuracy prevails as a defining element of this dimension.

---

[6] As one of the reviewers rightly observed, it is essential to clarify that my reference to diversity here pertains specifically to skills and career stages. The research community is not, however, necessarily diverse in terms of ethnicity, gender, or socioeconomic status.

In sum, this first dimension is shaped by the general inaccessibility of chatbot experimentation to wider audiences and the underlying force to expand the frontiers of conversational AI, continuously attaining the technical metrics and satisfying scientific inquiry. The commercial impact becomes more apparent in the dimension of *chatbots as commodities* that I will analyze in the ensuing section.

## 4.2  Chatbots as Commodities

While the first dimension is characterized by the technological advancement and development of these systems, which typically occurs in contexts restricted to the public and involves highly specialized teams, the second dimension refers to the circulation of chatbots in the market. Stated differently, the first dimension focuses on the inner technical enhancements and theoretical underpinnings of AI, while the second dimension acquires a business-oriented and for-profit nature. This commercial dimension involves the application or adaptation of conversational systems within a downstream framework, wherein chatbots transit through the market as commodities, being adopted by businesses of various kinds, aimed at leveraging these technologies to enhance customer engagement and streamline operations. In this dimension, I am using the term commodity in the basic form scholars from the field of economics employ it. If a commodity is a standardized good or service with a monetary value that traded through market conventions (Hashimzade et al., 2017; Law, 2018), then chatbots are indeed a commodities commercialized through the forces of the market.

The dimension of chatbots as marketed commodities emerged visibly during the 1990s. The milieu fostered by commercialization of personal computers, the widespread availability of the Internet and more robust datasets enabled the emergence of new chatbot models and lived experiences. The internet facilitated public access to chatbot demonstrations, allowing users to explore these systems firsthand. The increased exposure to chatbots helped to raise awareness of this technology and its potential applications across different market sectors. The 1990s are certainly a decade of relevance for the configuration of this dimension, as discourses around technology facilitating market transactions and economic growth resonated in both sides of the Atlantic (Bietti, 2023; Farrand, 2023).[7]

---

[7] Both Bietti and Farrand point out the role of regulation in the early days of widespread access to the Internet. Farrand contends that Europe, in the 1990s, embraced an ordoliberal approach that shaped the programmatic framework around opening telecommunications markets to competition and guarantees of universal service. In turn, Bietti employs a genealogical approach around platform regulation and distils three paradigmatic or archetypal conceptions of platform regulation: libertarian; liberal and neoliberal; critical approaches.

As chatbots gained commercial visibility, their integration across industries became more apparent. This represents a new form of organization and dynamics. As market trading becomes an allowed form of social organization and allocation of a given product or service, this logic is governed by supply and demand, pricing strategies, marketing, and valuation based on opportunity cost (Radin, 1987). In this realm, the pursuit of profit becomes the primary driving force behind the development of chatbot technologies. Although technological efficiency and innovation remain relevant, the focus predominantly leans towards enhancing market revenues. Consequently, instead of focusing on achieving AI milestones like passing the Turing Test,[8] this dimension places emphasis on developing profitable marketing approaches, commercializing the technology, and managing its distribution across markets.

The 1990s and early 2000s did not just witness the development of more sophisticated chatbots and greater accessibility; this era also marked the inception of the chatbot industry and the emergence of a new dimension. In other words, it was in the 1990s that chatbots left the walls of computers labs and universities and moved into the commercial sphere, becoming tradable and attractive systems for business and companies. This transition, which is underrepresented in scholarly discussions on conversational AI, marks what I label as the *chatbot commercial revolution*. This commercial dimension of chatbots laid the ground for dedicated specialized companies offering chatbot implementation services for customer service and marketing.[9] In this vein, a new market emerges when a convergence of demand, supply, and institutions interact to create a new space for economic transactions (Sarasvathy & Dew, 2005), and this second dimension of chatbot technology builds on that convergence.

Concurrently, this dimension is marked by the consolidation of major tech corporations, firstly technology powerhouses like IBM, Apple, and Microsoft, and later Google/Alphabet, Amazon, and Facebook/Meta, as well as their counterparts in China: Alibaba, Tencent, and Baidu. These entities not only dominate the market (Figueroa-Torres, 2023), but also mediate the information economy (Cohen, 2019). More particularly, with their established in-house AI research divisions and links with academic research units, Big Tech pushes forward their own research agendas in AI development—with generative models being certainly one of current priorities. In that regard, the emergence of this market dimension overlaps and affects the scientific dimension.

Meanwhile, the smaller businesses adopting and deploying chatbots, effectively become consumers of Big Tech innovations—what is typically termed in legal and economic discourse as business-to-business interactions. However, from a technological perspective, these smaller entities are essentially deployers or *intermediary users* of the infrastructures developed by larger firms, whereas the individual consumer who interacts the chatbot for commercial transactions constitutes the end-user, within a larger algorithmic supply chain.

---

[8] This test, in its roots, refers to making a human–machine interaction indistinguishable from a human–human one (Turing, 1950).

[9] It has to be noted that the academic literature over the history of commercial chatbots and providers is almost inexistent, however, repositories like chatbot.org are a useful tool to preserve and retrieve information about the evolution of the market of conversational agents, *See* (Verduijn, n.d.).

Figures from 2023, indicate that the integration of NLP and customer support systems resulted in a total market value worth of $8.1 billion—the use of chatbots is the most commonly adopted generative AI use case by function, followed by creating drafts of documents (Stanford University, 2024).

The market-driven dimension of chatbot technology is characterized by profit-oriented cycles, where commercial strategies focus on efficiency, profitability, and business alignment. In this dimension, chatbots are perceived not as mere computational innovations, but as tools adopted by businesses with the primary aim of optimizing operations and maximizing profit. This contrast does not suggest that market and scientific research are in conflict but highlights a distinct interplay of conditions and patterns that populate each dimension. Once this article distils the third and last dimension, it will explore how these dimensions overlap and interact, though it is important to acknowledge that despite their intersections, the underlying components remain different.

### 4.3  Chatbots as Intimate Companions

"Would you mind leaving the room, please?"—a phrase that portraits vividly Weizenbaum's secretary interacting with ELIZA, asking the computer scientist for privacy (Weizenbaum, 1967). From the early days of computational linguistics, chatbots afforded emotional attachments. What seems like a contemporary phenomenon finds echo in Weinbaum's days of experimentation.

While the first dimension is rooted in technological advancement within computer labs, and the second the dimension deals with the deployment of chatbots in pre-existing business environments, where they are implemented to enhance consumer interactions; the third dimension emerges in the personal use of chatbots, targeting the human needs for companionship and emotional interaction.

More than three decades ago, the paradigm of *computers as social actors* revealed that users often exhibit behaviors toward computers that defy logic when applied to inanimate objects but make perfect sense when directed at fellow humans—especially when these systems mimic human traits such as voice or text-based conversation (Nass et al., 1994). This behavior unsettles traditional expectation of artificial agents — placing them instead within a framework of social norms and emotional engagement. In recent years, this trend has intensified with chatbots that engage on the basis of specific emotional or sentimental needs, moving away from purely task-oriented functions. This engagement leads to emotional attachment (Sharpe & Ciriello, 2024), along with risks associated to their use (Boine, 2023). Frequently, while not exclusively, termed *social chatbots* in computer science literature (Gao et al., 2019; Shum et al., 2018), or Social AI (Sætra, 2020), these systems are crafted to satisfy emotional needs rather than functional tasks such as customer service or assistance.[10] Shevlin observes this wave of more accessible conversational systems acquiring significant but less visible roles in human lives, opting for the term of *SocialAI* (Shevlin, 2024). In this vein, I have referred elsewhere to this phenomenon as *affection as a service*, to highlight the for-profit logic and the dynamics of providers offering these emotional services (Figueroa-Torres, 2024).

Representative examples of chatbots used to fulfil human needs for connection and identity through digital means include AI companions in both hemispheres, like Replika (Pentina et al., 2023; Sharpe & Ciriello, 2024) and XiaoIce (Liberati, 2023).

The adoption of Replika across Western markets and XiaoIce in China, particularly through overreaching platforms like WeChat, QQ, and Weibo, highlights how users are turning to conversational agents to facilitate a sense of emotional connectivity and affection. Moreover, Liberati's insights into the Japanese context — with its unique clash between the notions of riajuu (the real) and otaku (imaginaries) — reveals how products have emerge to bridge this gap, like Love Plus [ラブプラス] and Gatebox [トップページ] offering the presence of a digital female character as a partner that proposes shared experiences such as dining, movie-going, and other social activities, embedding themselves into the user's daily life (Liberati, 2023).[11]

More recently, the inception of *MyAI* on Snapchat (Snapchat, 2023), powered by a customized implementation of OpenAI's ChatGPT, further exemplifies the current trend of deploying chatbots to offer users emotional engagement, targeting younger demographics with an in-platform bot that facilitating spontaneous and casual AI conversations into daily social media interactions. The case of Snapchat also places these interactions within the same digital spaces where human conversations occur, blurring traditional distinctions between user-to-user and user-to-machine communications.

Moreover, Meta has recently rolled out a beta release of 28 AI Avatars, each with distinct personalities and interests, incorporating the facial features of celebrities such as Snoop Dogg, Paris Hilton, and Tom Brady (Meta, 2023), enabling users to interact with virtual likenesses of figures they admire or follow, thereby personalizing the digital experience bridging the experiential gap between celebrity and fan.

Concurrently, in this dimension controversial developments have flourished, such as Janitor AI's sexbots (*Janitor AI - Terms of use*, 2023), which serve as a platform for NSFW chats, allowing users to explore their sexual desires. Moreover, the development of post-mortem chatbots (Figueroa-Torres, 2024; Harbinja, 2022; Hollanek & Nowaczyk-Basińska, 2024; Lindemann, 2022), which aim to digitally reincarnate deceased relatives, further illustrates how chatbots are expanding into emotionally complex domains like bereavement, and the imaginaries of digital reincarnation. In 2022, I coined the term chatbots to refer to these systems (Figueroa-Torres, 2022), but other notions like *griefbots*, *deadbots* or *deathbots* are also used in scholarly contributions.

---

[10] Sætra uses the term to encompass both digital interfaces and embodied artificial agents like robots. My notion of *affection as a service* is, for now, mostly concerned with conversational AI systems as digital interfaces.

[11] Another pressing area of inquiry in the chatbot dimension of *affection as a service* is how these are highly gendered, which had led to scholars to stress the inequities caused by technology's troubling relationship with gender, wherein a parallelism with game design becomes apparent: "In game design and research, feminist concerns have long been at the forefront of the field […] The portrayal of the female body, for example, is a commonly discussed topic in video game and virtual world research […] Such features embody a "male gaze," which makes all players spectators of the female body […] the construction and presentation of gender is very much part of the design of gaming experience" (Bardzell, 2010). These gendered constructs are present as well in virtual avatars or companions, as the unbalanced number of female vs male characters in providing sites may suggest.

Privacy concerns are particularly present in this dimension. A study by the Mozilla Foundation on 11 *romantic* chatbots revealed these systems collect extensive personal and sensitive information, including health details like sexual health and medication usage, and how these services lack transparency in their operations and data protection (Stanford University, 2024).

Digital friends, AI companions, ghostbots, sexbots, celebrity avatars are part of the constellation of this new dimension. Nonetheless, it is essential to raise awareness on the inaccurate differentiation of *commercial* versus *social* chatbots, as such distinction might oversimplify their roles. This is because despite their emotional engagement focus, social chatbots remain embedded within commercial frameworks. Thus, they are commercial, but their development reflects a broader aim to satisfy users' desires for communication, affection, and social belonging, transcending mere computational efficiency.

In addition to not being concerned with computational efficiency, the impetus here is not focused on facilitating commercial transactions, as in the commercial dimension.[12] Rather, the driving force in this dimension is the emotional component of chatbots.

Some of the literature on AI companions (Chaturvedi et al., 2023) proves informative when providing a thorough overview from Tamagotchi to Replika, and beyond. Yet, such a view overlooks a crucial shift: the difference between anthropomorphizing transactional chatbots and those designed specifically as social companions. Users may certainly project emotional characteristics onto transactional systems like customer service chatbots, but these reactions are incidental to the system's intended function—much like Weizenbaum's secretary did with ELIZA, as previously analyzed. In contrast, chatbots crafted specifically for intimate or personal interactions and attached to specific business models, such as those simulating romantic relationships or the personalities of deceased relatives, represent a fundamentally different phenomenon.

Acknowledging the emotional component — the inherent drive to interact with the bot on an emotional level — is embedded at the core of SocialAI systems is crucial for understanding the nature of this third dimension of chatbot technology. Intimate engagement is not an incidental aspect but a core feature.[13]

---

[12] Naturally, there is a for-profit logic in the commercialization of affection as a service bots, but it is distinct from the second dimension. Here, the crux is on profiting from the overall user experience and the unique attachment that the user forms with the bot, rather than assisting customers or executing individual transactions within a pipelined process.

[13] This is a phenomenon closely related to what scholars in the realm of communication studies and social psychology refer to as *parasocial relationships*, wherein subjects develop a one-sided connection

The emotional attachment, whereby individuals turn to chatbots for companionship, secrecy, sexual fulfillment, or grief management, among other emotions, illustrates this third dimension that — while influenced by the technological and market dimensions — operates distinctly by facilitating users some degree of perceived wellness or satisfaction, even providing a sense of fulfilment of something hardly achievable or unattainable goal in the physical realm: talking with a celebrity, having an idealized sexual partner or talking with a deceased relative, among other experiences, all serving the purpose of synthetic emotional experiences.

Against this backdrop, it is valid and necessary to challenge that these interactions, though intended to mimic human connection, might lead to different concerns such as social isolation and a weakening of community bonds. But such a claim does not erode this dimension. In fact, this dimension is marked by concerns around individuals turning to digital interfaces, instead of human beings, for support and companionship, which could weaken social ties and community bonds—critical components of societal health and individual well-being. Thus, these critiques are in fact part of this dimension and shape the debate that separates it from the other two previously examined.

## 5  Understanding the three Dimensions and Their Implications

Chatbots manifest across these three societal dimensions, each with its own unique set of characteristics, dynamics, interests, and discursive practices. These interplays can be organized as follows (Table 1):

The first dimensions centers on technological enhancement and progression, driven by an almost dogmatic focus on advancing conversational AI and pushing the limits of what chatbots can achieve. Metrics, performance benchmarks, accuracy assessment, computational capacity, tokenization, vectorization, parsing, and so forth, populate this dimension and drive the ever-rising expectations set in the field. This first dimension is manifested in the jargon and discussions of executives and lead engineers from major companies with their own in-house development teams, especially during public announcements about new developments in their systems and future directions. As exemplified by one executive's recent statement "Even if he spends $50 billion a year, some breakthroughs for mankind are priceless." (Fortune.com, 2024). Furthermore, this dimension also serves as fertile ground where certain academic units and scholars in 'hard-core' computer science flourish. They employ jargon —not too far from that of the executive just mentioned— in their applications for funding and in the recruitment of doctoral researchers. This implies a narrative of inevitable progression and the need to attain advancement at any cost, as if it is our destiny as humanity.

---

Footnote 13 (continued)

with public personas, such as a TV star, and believe these belong to their social circle "in the same way they know and understand flesh and blood friends". (Perse & Rubin, 1989).

**Table 1** Characteristics of each dimension of chatbot technology

| Dimension | Focus | Environment | Key characteristics |
|---|---|---|---|
| Scientific Research | Technological innovation and advancement of conversational AI | Computational labs, research units | *Restricted access. Primarily interests of computer scientists and tech developers. Continuous testing and benchmarking, even with industry shifts towards Big Tech leadership* |
| Marketed Commodities | Commercial adoption and profit | Business and commercial settings | *Focus on enhancing customer engagement and streamlining operations. Integration across industries. Development driven by supply–demand dynamics and profitability. Potential replacement of human labor for chatbots in certain customer transactions* |
| Intimate companions | Emotional and social fulfillment | Personal use and intimate environments | *Designed to satisfy emotional needs and enhance personal interaction. Raises social and ethical concerns about human–machine relationships. Also referred to as "Affection as Service", "Social AI" or "Emotional AI"* |

The experiences in the second dimension are shaped by an eye toward adapting chatbot technology to diverse industries and operations, with a grounded interest in leveraging these technologies in hopes of enhancing a firm or business's revenue, streamlining operations, enhancing customer service, and ultimately boosting profitability. The discourse in this dimension deals with automating tasks, optimizing resource allocation, and adopting what seems to be innovative business solutions. In this second dimension, there is a persuasive trend and frenzy where seemingly diverse areas, industries, and professions are, for some reason, taking the risks of introducing chatbot technology. This holds even in commercial activities where chatbots were not originally needed, yet are being used to allegedly facilitate and enhance their operations—legal advice, medical diagnosis, accounting, or teaching— in addition to mainstream realms, like customer support, that fall prey to introducing iterations of chatbot technology without regard to the specific underlying technical functioning and the risks. A notable case is that of a commercial airline chatbot inventing a reimbursement policy in a customer conversation, resulting in the airline being sued in court and ultimately being force to honor the hallucinated policy (Cecco, 2024).

The umbrella terms of the third dimension are companionship and affection. This dimension shares similarities with the second dimension—chatbots as strategies deployed by companies to facilitate transactions—where certain business practices are increasingly mediated by chatbots. Similarly, in this third dimension, specific aspects of human life are becoming *chatbotified*. However, the outcomes differ significantly. In the commercial realm, the primary focus is on the profit of the company deploying the chatbot within established commercial practices. In contrast, in this third dimension, it is the user who seeks to fulfill emotional needs through these interactions. This third dimension highlights concerns and legitimate doubts on to what extent we are becoming dependent on these systems in our private lives. Such concerns have sparked in both sides of the Atlantic as chatbot-related suicide instances have been reported. A man in Belgium apparently committed suicide after interactions with a chatbot (Xiang, 2023). Similarly, a bereaved mother in Florida is suing in court a chatbot company, as she contends that her teenage son committed suicide after weeks of attachment to an AI companion (Associated Press, 2024).

## 6 Overlapping Dimensions: Could There Be a Fourth Dimension?

In structuring our understanding of chatbot technology, as highlighted in the preceding sections, this article opted for a dimensional framework that emphasizes the overlapping and interconnected dimensions to understand, analyze, and critique chatbot technology. Take, for example, the clear overlap between the dimensions of scientific research and the market, with Big Tech being the main sponsor of AI research. But in addition to that, the market might well fuel specific chatbot models that profit from emotional needs, leading to the commercialization of systems designed to fulfill these desires—something that can be placed within the wider phenomenon that has been termed *emotional capitalism* (Illouz, 2009). In turn, developers, often motivated by personal narratives like grief, loneliness, or even sexual dissatisfaction, might pursue specific models and interfaces. Take the case of Replika as an illustrative example (Pentina et al., 2023), it began as a computational experiment out of grief and now operates within a commercial sphere, providing virtual friends and avatars.[14]

Similarly, chatbots addressing intimate aspects of life, such as faith, reveal another interesting overlap. *Habibi.AI* (*Habibi AI*, 2024), designed for the Muslim community, and *Text With Jesus* (*Text With Jesus—AI Chatbot App*, 2024), a subscription-based service, illustrate how scientific research and intimate dimensions entangle and influence technological development. In that regard, conversational systems might well be the result of developers driven by personal motivations and leveraging their technical skills—demonstrating the complex, overlapping nature of chatbot dimensions.

But in outlining this dimensional framework, it is important to acknowledge that this article does not claim to have exhausted the possibilities. While the focus here is on three identified dimensions, the door remains open for a fourth dimension — or perhaps even more — to emerge. Though it would be premature to delineate a fully realized fourth dimension, speculative academic insight may serve as a platform to imagine possible scenarios. For instance, scholarly contributions have explored settings where chatbots may take on semi-public or semi-authoritative roles, such as serving as judges or mediators in dispute resolution (Shonk, 2024) or interpreting legal documents like wills (Schafer, 2020). Such roles would transcend the private, commercial, and emotional dimensions, influencing not just the user experience but also the actual distribution of resources and legal outcomes. In this same line, for example, the potential use of chatbots to manage public services like health, taxation, or education (Androutsopoulou et al., 2019) warrants further scrutiny—as the end-user must not be seen as a customer but as a citizen. These are speculative settings, but they may suggest an interplay of dynamics, patterns of influence, interconnected set of structures, conditions, experiences, and narratives where chatbot technology could transcend into a new dimension.

There may well be further dimensions, shaped by different defining elements and emerging as chatbots continue to penetrate broader areas of society. This three-dimensional framework, by its very nature, is not exhaustive, and it precisely acknowledges the complexity of digital technologies and their applications in ways that transcend simple linear progression or chronological development.

---

[14] For two years, Eugenia Kuyda —the founder of Replika— had been building Luka, a bot-based messaging app, initially designed for restaurant reservations. After a close friend died, friends discussed ways to honor his memory—ideas ranged from a coffee-table book of his life to a memorial website. Yet, to Kuyda, all these felt insufficient. Grieving, she reread the thousands of texts Mazurenko had sent her, finding comfort in them. She first did a chatbot emulating her deceased friend. By 2016, Kuyda had shifted Luka's focus from restaurant bots to developing emotionally resonant chatbots, culminating in a project called Replika— leading to the for-profit company of the same name. (Newton, 2016).

# 7 Concluding Remarks

This article organized a dimensional framework for understanding and analyzing the evolution of chatbot technology. Each of the three dimensions corresponds to distinct dynamics, patterns of influence, interconnected set of structures, and defining elements that shape how chatbots, as conversational systems, inhabit different social spheres and impact lived experiences. The first dimension addresses the technological advancements of conversational AI, where chatbots are computational experiments subject to continuous optimization in capabilities and performance. In the second dimension chatbots are elements of the marketplace, where they function as commercialized tools designed to optimize services, maximize profits, and replace human labor. In the third dimension, chatbots become intimate companions, systems used for satisfying emotional needs such as friendship, sexual desire, or grief, thereby integrating into deeply personal aspects of life.

The framework of chatbots through three dimensions—from labs, the market, and the bedroom—also underscores the open-ended nature of chatbot development, allowing for future reflection to explore the emergence of further dimensions. Ideally, this framework invites deeper scholarly engagement. In the end, while we *prompt* chatbots to execute diverse tasks, chatbot technology, in turn, *prompts* us to think beyond its initial technical configurations and consider the far-reaching implications of its ever-evolving nature.

**Acknowledgements** I appreciate the peer reviewers' time and constructive comments. Thanks to Joshua Jowitt, Gaurav Kudtarkar, Ilke Turkmendag, Tina Sikka, Divij Joshi, and Benjamin Farramd for their helpful comments on different drafts of this paper. Additionally, I'd like to express my gratitude to the audiences at *Open Lab*, Newcastle School of Computing, UK; the *University of Delhi* and *Jindal Global University*, India; and the *University of Buenos Aires*, Argentina, for their engagement and insightful discussions during the guest lectures and seminars where I advanced and discussed some of the ideas part of this paper.

**Authors' contributions** Not applicable.

**Funding** The author has received research funding by the UK Arts and Humanities Research Council, thorough the Northern Bridge Consortium, Gran Number 3000024088.

**Data Availability** Not applicable.

## Declarations

# References

AbuShawar, B., & Atwell, E. (2015). ALICE chatbot: Trials and outputs. *Computación y Sistemas, 19*(4), 625–632.

Adamopoulou, E., & Moussiades, L. (2020). Chatbots: History, technology, and applications. *Machine Learning with Applications, 2*, 100006. https://doi.org/10.1016/j.mlwa.2020.100006

Androutsopoulou, A., Karacapilidis, N., Loukis, E., & Charalabidis, Y. (2019). Transforming the communication between citizens and government through AI-guided chatbots. *Government Information Quarterly, 36*(2), 358–367. https://doi.org/10.1016/j.giq.2018.10.001

Associated Press. (2024, October 25). An AI chatbot pushed a teen to kill himself, a lawsuit against its creator alleges. *AP News*. https://apnews.com/article/chatbot-ai-lawsuit-suicide-teen-artificial-intelligence-9d48adc572100822fdbc3c90d1456bd0

Bardzell, S. (2010). Feminist HCI: Taking stock and outlining an agenda for design. *Proceedings of the SIGCHI Conference on Human Factors in Computing Systems*, 1301–1310. https://doi.org/10.1145/1753326.1753521

Bender, E. M., Gebru, T., McMillan-Major, A., & Shmitchell, S. (2021). On the Dangers of Stochastic Parrots: Can Language Models Be Too Big?. *Proceedings of the 2021 ACM Conference on Fairness, Accountability, and Transparency*, 610–623. https://doi.org/10.1145/3442188.3445922

Bietti, E. (2023). A genealogy of digital platform regulation. *Geo. l. Tech. Rev., 7*, 1.

Birnhack, M. D., & Elkin-Koren, N. (2003). The invisible handshake: The reemergence of the state in the digital environment. *Va. JL & Tech., 8*, 1.

Boine, C. (2023). Emotional Attachment to AI Companions and European Law. *MIT Case Studies in Social and Ethical Responsibilities of Computing*, *Winter 2023*. https://doi.org/10.21428/2c646de5.db67ec7f

Brown, T. B., Mann, B., Ryder, N., Subbiah, M., Kaplan, J., Dhariwal, P., Neelakantan, A., Shyam, P., Sastry, G., Askell, A., Agarwal, S., Herbert-Voss, A., Krueger, G., Henighan, T., Child, R., Ramesh, A., Ziegler, D. M., Wu, J., Winter, C., … Amodei, D. (2020). *Language Models are Few-Shot Learners* (arXiv:2005.14165). arXiv. https://doi.org/10.48550/arXiv.2005.14165

Cecco, L. (2024, February 16). Air Canada ordered to pay customer who was misled by airline's chatbot. *The Guardian*. https://www.theguardian.com/world/2024/feb/16/air-canada-chatbot-lawsuit

Center for Research on Foundation Models. (2021). *On the Opportunities and Risks of Foundation Models*. https://fsi.stanford.edu/publication/opportunities-and-risks-foundation-models

Chaturvedi, R., Verma, S., Das, R., & Dwivedi, Y. K. (2023). Social companionship with artificial intelligence: Recent trends and future avenues. *Technological Forecasting and Social Change, 193*, 122634. https://doi.org/10.1016/j.techfore.2023.122634

Cohen, J. E. (2017). Affording fundamental rights: A provocation inspired by Mireille Hildebrandt. *Critical Analysis of Law, 4*, 78.

Cohen, J. E. (2019). *Between truth and power: The legal constructions of informational capitalism*. Oxford University Press.

*CUI Conference—Proceedings*. (2024). ACM Digital Library. https://dl.acm.org/conference/cui/proceedings

Edwards, P. N. (1996). *The closed world: Computers and the politics of discourse in Cold War America*. MIT Press.

Farrand, B. (2023). The ordoliberal internet? Continuity and change in the EU's approach to the governance of cyberspace. *European Law Open*, *2*(1), 106–127. Cambridge Core. https://doi.org/10.1017/elo.2023.14

Feenberg, A. (1999). *Questioning technology*. Routledge.

Figueroa-Torres, M. (2022, January 18). *Ghostbots, the quest for digital immortality and the law*. https://www.jurist.org/commentary/2022/01/mauricio-figueroa-ghostbots-digital-immortality-law/

Figueroa-Torres, M. (2023). Big Tech Platforms, Democracy and the Law: Global Problems, Legal Perspectives and the Mexican Experience. *Mexican Law Review*, *15*(2), 3–22. https://doi.org/10.22201/iij.24485306e.2023.2.17615

Figueroa-Torres, M. (2024). Affection as a service: Ghostbots and the changing nature of mourning. *Computer Law & Security Review, 52*, 105943. https://doi.org/10.1016/j.clsr.2024.105943

Floridi, L. (2014). *The fourth revolution: How the infosphere is reshaping human reality*. OUP Oxford.

Floridi, L. (2023). On Good and Evil, the Mistaken Idea That Technology Is Ever Neutral, and the Importance of the Double-Charge Thesis. *Philosophy & Technology*, *36*(3), 60, s13347–023–00661–00664. https://doi.org/10.1007/s13347-023-00661-4

Fortune.com. (2024). *OpenAI's Sam Altman doesn't care how much AGI will cost: Even if he spends $50 billion a year, some breakthroughs for mankind are priceless*. Fortune. https://fortune.com/2024/05/03/openai-sam-altman-microsoft-agi-artificial-general-intelligence-costs/

Fryer, L., Coniam, D., Carpenter, R., & Lăpușneanu, D. (2020). *Bots for language learning now: Current and future directions*. https://scholarspace.manoa.hawaii.edu/handle/10125/44719

Gao, J., Galley, M., & Li, L. (2019). Neural Approaches to Conversational AI. *Foundations and Trends® in Information Retrieval*, *13*(2–3), 127–298. https://doi.org/10.1561/1500000074

*Habibi AI*. (2024). https://askhabibi.ai

Harbinja, E. (2022). *Digital death, digital assets and post-mortem privacy: Theory, technology and the law*. Edinburgh University Press. https://doi.org/10.1515/9781474485388

Hashimzade, N., Myles, G., & Black, J. (2017). Commodity market. *Oxford University Press*. https://doi.org/10.1093/acref/9780198759430.013.0463

Hollanek, T., & Nowaczyk-Basińska, K. (2024). Griefbots, Deadbots, Postmortem Avatars: On Responsible Applications of Generative AI in the Digital Afterlife Industry. *Philosophy & Technology, 37*(2), 63. https://doi.org/10.1007/s13347-024-00744-w

Illouz, E. (2009). Emotions, Imagination and Consumption: A new research agenda. *Journal of Consumer Culture, 9*(3), 377–413. https://doi.org/10.1177/1469540509342053

*Janitor AI - Terms of use*. (2023). JanitorAI. https://janitorai.com/term

Latour, B. (1997). *Science in action: How to follow scientists and engineers through society* (7th printing..). Harvard University Press.

Law, J. (2018). Commodity. Oxford University Presss. https://doi.org/10.1093/acref/9780198789741.013.0710

Liberati, N. (2023). Digital Intimacy in China and Japan. *Human Studies, 46*(3), 389–403. https://doi.org/10.1007/s10746-022-09631-9

Lindemann, N. F. (2022). The Ethics of 'Deathbots.' *Science and Engineering Ethics, 28*(6), 60. https://doi.org/10.1007/s11948-022-00417-x

Mauldin, M. L. (1994). Chatterbots, tinymuds, and the turing test: Entering the loebner prize competition. *AAAI*, *94*, 16–21. https://cdn.aaai.org/AAAI/1994/AAAI94-003.pdf

McTear, M. (2020). *Conversational AI: Dialogue systems, conversational agents, and chatbots*. Morgan & Claypool Publishers. https://ieeexplore.ieee.org/servlet/opac?bknumber=9299458

Meta. (2023, September 27). Introducing new AI experiences across our family of apps and devices. *Meta*. https://about.fb.com/news/2023/09/introducing-ai-powered-assistants-characters-and-creative-tools/

Nass, C., Steuer, J., & Tauber, E. R. (1994). Computers are social actors. *Proceedings of the SIGCHI Conference on Human Factors in Computing Systems Celebrating Interdependence - CHI '94*, 72–78. https://doi.org/10.1145/191666.191703

NeurIPS. (2024). *Neural information processing systems archive*. https://neurips.cc/FAQ

Newton, C. (2016, October 6). *When her best friend died, she used artificial intelligence to keep talking to him*. TheVerge.Com. http://www.theverge.com/a/luka-artificial-intelligence-memorial-roman-mazurenko-bot

OpenAI. (2020). *GPT-3 Model Card*. GitHub. https://github.com/openai/gpt-3/blob/master/model-card.md

OpenAI. (2022). *Introducing ChatGPT*. https://openai.com/blog/chatgpt

Pentina, I., Hancock, T., & Xie, T. (2023). Exploring relationship development with social chatbots: A mixed-method study of replika. *Computers in Human Behavior, 140*, 107600.

Ⓢ Springer

Perez, C. (2010). Technological revolutions and techno-economic paradigms. *Cambridge Journal of Economics, 34*(1), 185–202.

Perse, E. M., & Rubin, R. B. (1989). Attribution in Social and Parasocial Relationships. *Communication Research, 16*(1), 59–77. https://doi.org/10.1177/009365089016001003

Pinch, T. J., & Bijker, W. E. (1984). The Social Construction of Facts and Artefacts: Or How the Sociology of Science and the Sociology of Technology might Benefit Each Other. *Social Studies of Science, 14*(3), 399–441. https://doi.org/10.1177/030631284014003004

Radin, M. J. (1987). Market-Inalienability. *Harvard Law Review*, 1849–1937.

Sætra, H. S. (2020). The Parasitic Nature of Social AI: Sharing Minds with the Mindless. *Integrative Psychological & Behavioral Science*, *54*(2).

Sarasvathy, S. D., & Dew, N. (2005). New market creation through transformation. *Journal of Evolutionary Economics, 15*(5), 533–565. https://doi.org/10.1007/s00191-005-0264-x

Schafer, B. (2020). On living and undead Wills: ZombAIs, technology and the future of inheritance law. In *Future law: Emerging technology, regulation and ethics* (pp. 225–261). Edinburgh University Press.

Schuchmann, S. (2019). *Analyzing the Prospect of an Approaching AI Winter.* https://doi.org/10.13140/RG.2.2.10932.91524

Sharpe, P., & Ciriello, R. F. (2024). Exploring attachment and trust in AI companion use. In *Australasian conference on information systems.* https://www.researchgate.net/profile/Raffaele-Ciriello/publication/385349277_Exploring_Attachment_and_Trust_in_AI_Companion_Use/links/67211b9977b63d1220cb621f/Exploring-Attachment-and-Trust-in-AI-Companion-Use.pdf

Shevlin, H. (2024). All too human? Identifying and mitigating ethical risks of Social AI. *Law, Ethics & Technology*, *1*(2). https://doi.org/10.55092/let20240003

Shonk, K. (2024, June 17). *AI mediation: Using AI to help mediate disputes.* PON - Program on Negotiation at Harvard Law School. https://www.pon.harvard.edu/daily/mediation/ai-mediation-using-ai-to-help-mediate-disputes/

Shum, H.-Y., He, X., & Li, D. (2018). From Eliza to XiaoIce: Challenges and opportunities with social chatbots. *Frontiers of Information Technology & Electronic Engineering, 19*, 10–26.

Snapchat. (2023). *What is my AI on Snapchat, and how do I use it?* Snapchat Support. https://help.snapchat.com/hc/en-gb/articles/13266788358932-What-is-My-AI-on-Snapchat-and-how-do-I-use-it-

Stanford University. (2024). *AI Index Report 2024 – Artificial Intelligence Index* (p. 502). Institute for Human-Centered AI. https://aiindex.stanford.edu/report/

Stanford University, School of Humanities and Sciences. (2024). *Program in STS.* https://sts.stanford.edu/

Tarleton Gillespie. (2016). Algorithm. In Benjamin Peters (Ed.), *Digital Keywords* (pp. 18–30). Princeton University Press. https://doi.org/10.2307/j.ctvct0023.6

*Text With Jesus—AI Chatbot App.* (2024). https://textwith.me/jesus/

Turing, A. M. (1950). Computing Machinery and Intelligence. *Mind, LIX*(236), 433–460. https://doi.org/10.1093/mind/LIX.236.433

UCL. (2024). *Science and Technology Studies.* Science and Technology Studies. https://www.ucl.ac.uk/sts

Verbeek, P.-P. (2005). *What Things Do: Philosophical Reflections on Technology, Agency, and Design.* Penn State Press.

Verduijn, X. (n.d.). *Chatbot Verbot, Verbots | Virtual Assistant Verbot | Virtual agent Verbot | Chat bot Verbot | Conversational agent Verbot | (4040).* Chatbots.Org. Retrieved June 8, 2023, from https://www.chatbots.org/chatterbot/verbot/

Vinyals, O., & Le, Q. (2015). *A neural conversational model* (arXiv:1506.05869). arXiv. http://arxiv.org/abs/1506.05869

Weizenbaum, J. (1966). ELIZA—a computer program for the study of natural language communication between man and machine. *Communications of the ACM, 9*(1), 36–45.

Weizenbaum, J. (1967). Contextual understanding by computers. *Communications of the ACM, 10*(8), 474–480. https://doi.org/10.1145/363534.363545

Weizenbaum, J. (1976). *Computer power and human reason: From judgment to calculation* (pp. xii, 300). W. H. Freeman & Co.

Xiang, C. (2023, March 30). "He would still be here": Man dies by suicide after talking with AI chatbot, widow says. *Vice.* https://www.vice.com/en/article/pkadgm/man-dies-by-suicide-after-talking-with-ai-chatbot-widow-says

**Publisher's Note**  Springer Nature remains neutral with regard to jurisdictional claims in published maps and institutional affiliations.